\documentclass[manuscript,screen]{acmart}
\renewcommand\footnotetextcopyrightpermission[1]{} 
\AtBeginDocument{%
  \providecommand\BibTeX{{%
    \normalfont B\kern-0.5em{\scshape i\kern-0.25em b}\kern-0.8em\TeX}}}

\usepackage{multirow,makecell}
\usepackage{tcolorbox}
\usepackage{color,xcolor}
\usepackage{listings,amsfonts}
\usepackage{caption}
\usepackage{subcaption}
\usepackage{threeparttable}
\usepackage{bbding}
\usepackage{graphicx}
\usepackage{booktabs} 
\usepackage{longtable}
\usepackage{array}
\usepackage{url}

\AtEndPreamble{
	\usepackage{hyperref}
	\hypersetup{
		colorlinks = true,
		linkcolor = purple,
		anchorcolor = purple,
		citecolor = purple,
		filecolor = purple,
		urlcolor = purple
	}
}

\usepackage{tcolorbox}
\def\BibTeX{{\rm B\kern-.05em{\sc i\kern-.025em b}\kern-.08em
    T\kern-.1667em\lower.7ex\hbox{E}\kern-.125emX}}

\newcommand{\ea}{et~al.}

\begin{document}

\title{LLM App Store Analysis: A Vision and Roadmap}

\author{Yanjie Zhao}
\email{yanjie_zhao@hust.edu.cn}
\authornotemark[2]
\affiliation{%
  \institution{Huazhong University of Science and Technology}
  \city{Wuhan}           
  \country{China}
}
\author{Xinyi Hou}
\email{xinyihou@hust.edu.cn}
\authornotemark[2]
\affiliation{%
  \institution{Huazhong University of Science and Technology}
  \city{Wuhan}           
  \country{China}
}
\author{Shenao Wang}
\email{shenaowang@hust.edu.cn}
\authornotemark[2]
\affiliation{%
  \institution{Huazhong University of Science and Technology}
  \city{Wuhan}
  \country{China}
}
\author{Haoyu Wang}
\authornote{Haoyu Wang is the corresponding author (haoyuwang@hust.edu.cn).}
\authornote{The full name of the author's affiliation is Hubei Key Laboratory of Distributed System Security, Hubei Engineering Research Center on Big Data Security, School of Cyber Science and Engineering, Huazhong University of Science and Technology.}
\email{haoyuwang@hust.edu.cn}
\affiliation{%
  \institution{Huazhong University of Science and Technology}
  \city{Wuhan}
  \country{China}
}

\begin{abstract}

The rapid growth and popularity of large language model (LLM) app stores have created new opportunities and challenges for researchers, developers, users, and app store managers. As the LLM app ecosystem continues to evolve, it is crucial to understand the current landscape and identify potential areas for future research and development. This paper presents a forward-looking analysis of LLM app stores, focusing on key aspects such as data mining, security risk identification, development assistance, and market dynamics. Our comprehensive examination extends to the intricate relationships between various stakeholders and the technological advancements driving the ecosystem's growth. We explore the ethical considerations and potential societal impacts of widespread LLM app adoption, highlighting the need for responsible innovation and governance frameworks. By examining these aspects, we aim to provide a vision for future research directions and highlight the importance of collaboration among stakeholders to address the challenges and opportunities within the LLM app ecosystem. The insights and recommendations provided in this paper serve as a foundation for driving innovation, ensuring responsible development, and creating a thriving, user-centric LLM app landscape.

\end{abstract}

\maketitle

\section{Introduction}

Large language models (LLMs), like GPT-4~\cite{achiam2023gpt} and LLaMA~\cite{touvron2023llama}, have significantly advanced the field of natural language processing. Trained on extensive textual datasets, these models excel in understanding complex language nuances and performing a variety of tasks with high proficiency. Applications of LLMs span generating human-like text, answering complex queries, and supporting functionalities in chatbots, content creation, language translation, and sentiment analysis. Their versatility has attracted considerable interest from various industries, with both established technology companies and startups aiming to harness these models for practical solutions~\cite{geminiteam2024geminifamilyhighlycapable,bai2023qwentechnicalreport}. Ongoing research efforts have led to rapid improvements in LLM performance, efficiency, and user accessibility~\cite{hou2023large,thirunavukarasu2023large}, thereby increasing the demand for LLM-driven applications and highlighting the necessity for more accessible platforms to deploy these advanced AI technologies for real-world scenarios.

The growing interest in LLMs has given rise to a new ecosystem: \textbf{LLM app stores}. Platforms such as OpenAI's GPT Store~\cite{gptstore}, Quora's Poe~\cite{poeexplore}, ByteDance's Coze~\cite{coze}, and FlowGPT~\cite{flowgpt} serve as centralized marketplaces for applications (apps) powered by LLMs. These app stores function as intermediaries between developers who create innovative LLM-based solutions and users seeking to utilize AI capabilities in personal or professional contexts. By offering a curated selection of LLM apps, these platforms democratize access to advanced AI technologies, extending their availability beyond AI researchers and technically proficient individuals. Users can explore and engage with a diverse array of LLM-powered apps across domains including productivity, education, entertainment, and personal assistance. This increased accessibility is fostering wider adoption of AI technologies, enabling both individuals and businesses to leverage the potential of LLMs without requiring extensive technical knowledge.

The emergence of LLM app stores represents a significant development in the AI domain, altering the manner in which advanced language models are utilized and experienced. These platforms are not solely marketplaces but also act as incubators for innovation, providing developers with essential tools, infrastructure, and an audience to realize their concepts. By streamlining the process from idea to deployment, LLM app stores accelerate the advancement of AI innovation and app development. This ecosystem encourages competition among developers, leading to improvements in app quality, user experience, and the exploration of novel use cases. Moreover, the interactive feedback loop between users and developers facilitated by these platforms allows for rapid iteration and refinement based on practical usage and requirements. Consequently, there is a proliferation of creative apps leveraging LLMs in previously unexplored ways, such as personalized learning assistants~\cite{personalized-learning-assistant} and AI-driven creative writing tools~\cite{AIwriting}.  The integration of multimodal functionalities—enabling LLMs to process and generate not only text but also images, audio, and video—opens new avenues for innovative and practical apps. Industry-specific LLM apps are also being developed, addressing the specialized needs of sectors such as healthcare, finance, legal services, and retail. Using the example of IKEA GPT~\cite{IKEA}, enterprises can develop their LLM apps by providing a specialized knowledge base that enables users to efficiently access product information through interactive dialogue. By hosting such an app on the GPT Store, enterprises can circumvent the complexities associated with LLM infrastructure, minimize the need for customer service personnel, and reduce advertising and marketing expenditures.

Despite the benefits, the rise of LLM app stores introduces challenges that require careful consideration. As these platforms gain prominence, issues concerning data privacy, security, and the ethical use of AI become increasingly significant. There is a pressing need for robust mechanisms to ensure the quality and safety of apps distributed through these stores, safeguarding users from potential misuse or harmful applications of LLMs. Additionally, matters related to intellectual property rights, equitable compensation for developers, and the risk of monopolistic practices within the LLM app ecosystem necessitate thorough examination and regulation. Establishing standards and best practices for LLM app development and distribution is thus essential for the sustainable growth of this ecosystem.

These multifaceted challenges underscore the need for comprehensive research and analysis in this rapidly evolving field. However, current literature predominantly focuses on the technical aspects of LLMs themselves, \textbf{leaving a significant gap in understanding the ecosystem that has developed around them}. The absence of in-depth studies on LLM app stores, their market dynamics, security implications, and impact on user experience represents a critical oversight in the current body of knowledge. To fill the existing knowledge gap, this paper aims to provide a forward-looking analysis of LLM app stores. Our research focuses on key aspects that shape the user experience, developer strategies, and the dynamics of the ecosystem, offering insights that are crucial for navigating the complexities of this emerging field.
In summary, our main contributions are as follows:

\begin{itemize}
    \item We present a comprehensive analysis of existing mainstream LLM app stores, showcasing data on user engagement, app quantities, and other relevant metrics. This analysis provides insights into the current state of the LLM app ecosystem, offering a foundation for future research and development in this area.
    \item We propose a research roadmap that outlines key areas for investigation, including LLM app data collection, security and privacy analysis, and ecosystem and market dynamics. This roadmap serves as a guide for researchers and practitioners, highlighting critical areas that require attention and further study.
    \item We discuss the implications arising from the research on LLM app stores, analyze challenges faced by the ecosystem, and provide recommendations for LLM app store stakeholders. This discussion aims to foster a more secure, ethical, and user-centric development of the LLM app ecosystem.
\end{itemize}

By providing these insights and recommendations based on a comprehensive analysis of the current landscape, this paper contributes to the development of a thriving, responsible LLM app ecosystem. Our work serves as a crucial starting point for future research, highlighting the importance of collaboration among stakeholders in addressing challenges and leveraging opportunities presented by LLM app stores.

\section{Definitions and Motivations}
\label{sec:motivation}

\begin{figure*}[ht!]
    \centering
    \includegraphics[width=0.9\linewidth]{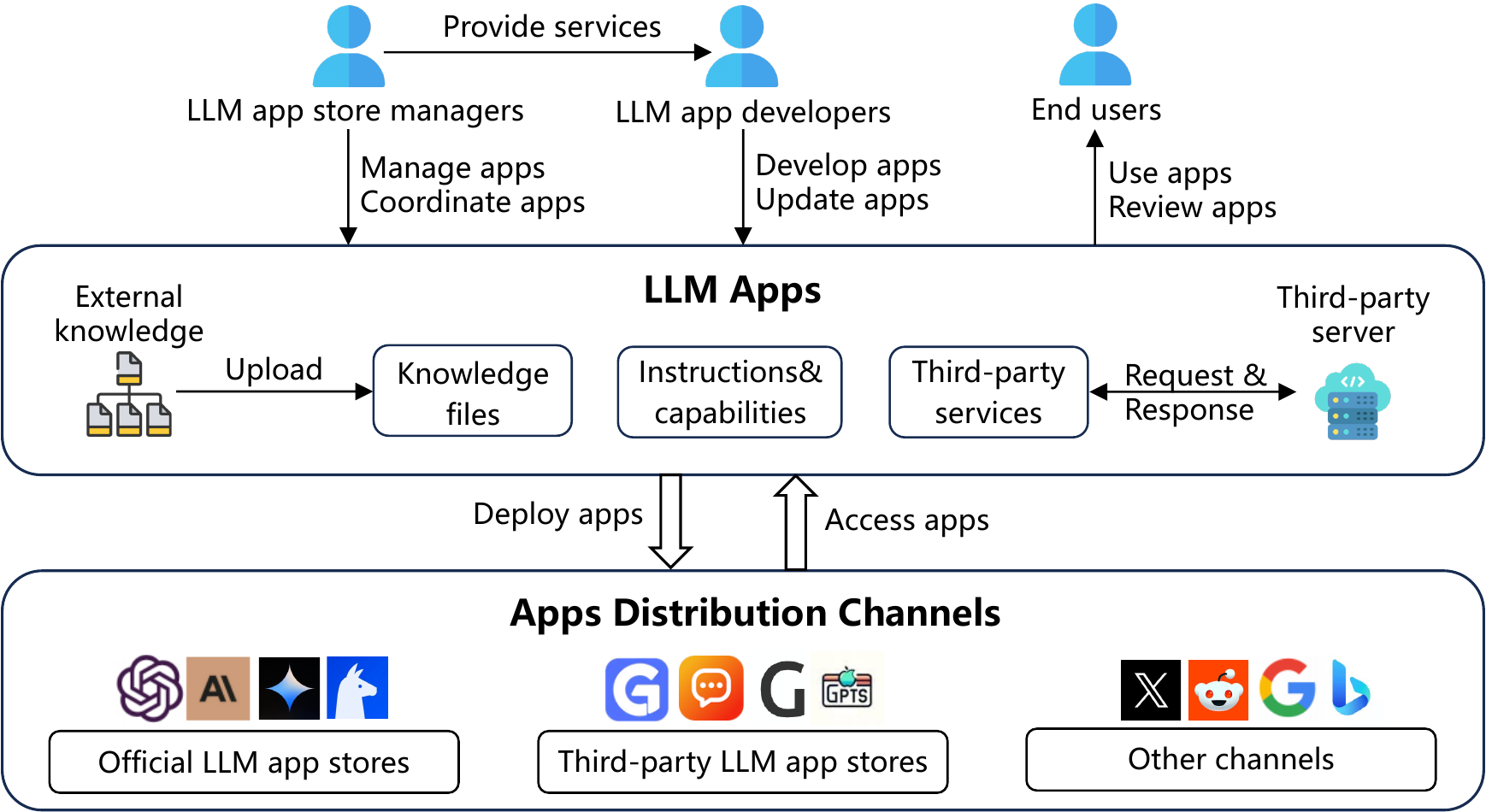}
	\caption{LLM app ecosystem components and operating mechanisms.}
    \label{fig:ecosystem}
\end{figure*}

\subsection{Definitions}

To provide the foundation for our analysis and discussions, it is essential to establish clear definitions of key concepts and terms related to \textit{LLM app} and \textit{LLM app store}. The following definitions explain these elements, serving as a reference point for the subsequent sections of this paper.

\begin{itemize}
    \item \textit{LLM app}: An \textit{LLM app} is a specialized application designed specifically to leverage the capabilities of LLMs, providing unique functionalities tailored to meet distinct tasks or user requirements. Unlike conventional \textit{mobile apps}, which may incorporate LLM technology as an additional feature, \textbf{LLM apps are fundamentally built to exploit the advanced processing power and cognitive capabilities of LLMs, relying heavily on platforms and hardware infrastructures typically provided by LLM providers}. These platforms—such as GPT Store, Poe, Coze, and FlowGPT—not only host the apps but also furnish the necessary computational resources and optimized environments essential for their operation. This reliance on powerful cloud-based hardware and specialized platforms distinguishes LLM apps from traditional mobile apps, which generally run locally on a user's device with additional external processing if necessary. LLM apps serve a wide range of purposes, from creative content generation and conversational agents to complex data analysis and problem-solving tasks. Their design prioritizes user experience, focusing on usability and efficiency to ensure that users can fully harness the sophisticated functionalities and adaptive nature of LLMs.

    \item \textit{LLM app store}: An \textit{LLM app store} is a centralized digital platform that serves as a repository for hosting, curating, and distributing LLM apps. It enables users to discover, evaluate, and access a variety of intelligent services tailored to their needs. These platforms not only facilitate the transaction and distribution process but also play a pivotal role in shaping the ecosystem by providing an organized environment where developers can showcase their creations while users can navigate through a plethora of options easily.
\end{itemize}

As illustrated in \autoref{fig:ecosystem}, the \textit{LLM app ecosystem} represents a collaborative environment that harnesses the transformative power of LLMs to develop specialized apps for a diverse user base. Within this ecosystem, \textit{LLM app store managers} are instrumental in enhancing the visibility and accessibility of LLM apps. They assist \textit{LLM app developers} by providing essential resources such as comprehensive documentation, technical support, and marketing assistance, thereby facilitating the development and launch of innovative LLM apps. Furthermore, they focus on ensuring a seamless user experience for \textit{end-users} by incorporating intuitive search and navigation features, enabling users to efficiently find the LLM apps that best align with their needs. These managers also streamline the transaction processes, creating pathways for developers to monetize their innovations effectively.

\textit{LLM app developers (creators)} represent the core of the ecosystem's innovation landscape. These individuals or teams are responsible for creating and customizing LLM apps to meet specific requirements and use cases. During the development phase, they carefully design the app's instructions and define its desired capabilities, which may include functionalities such as web browsing, image generation, or code interpretation. Developers can further augment their apps by integrating external knowledge sources or connecting to third-party services through established protocols such as API keys or OAuth mechanisms~\cite{leiba2012oauth}. This ability to enrich the apps not only enhances their functionality but also broadens their versatility in addressing user demands.
Once an LLM app has been developed, the creators can deploy and submit their apps to various distribution channels. These channels encompass \textit{official LLM app stores}, \textit{third-party LLM app stores}, and popular social media platforms like X~\cite{xcorp} and Reddit~\cite{reddit}, as well as search engines such as Google~\cite{google} and Bing~\cite{bing}. This multifaceted approach to distribution ensures that the apps achieve maximum visibility and reach, thereby fostering greater discoverability among potential users.

\textit{End-users}, which include individuals, businesses, or organizations, constitute the consumer base of the LLM app ecosystem. They engage with the platform by browsing and discovering the available LLM apps through diverse distribution channels. End-users have the option to purchase or acquire these apps, as well as provide reviews and feedback, which is invaluable for ongoing improvements and iterations within the ecosystem. Their experiences and insights contribute to shaping the development and refinement of future LLM apps, creating a dynamic feedback loop that benefits all stakeholders involved.

\begin{table}[ht!]
    \centering
    \caption{Prominent official LLM app stores and their associated metrics.}
    \resizebox{\linewidth}{!}{%
    \begin{tabular}{lp{0.2\linewidth}p{0.2\linewidth}p{0.2\linewidth}p{0.22\linewidth}}
    \toprule[1.2pt]
        \textbf{LLM App Store} & \textbf{Affiliation} & \textbf{Portal} & \textbf{Popularity} & \textbf{Platform} \\
        \hline
        GPT Store & OpenAI & \url{https://chatgpt.com/gpts} & $\geq 3,000,000$ GPTs \cite{openai2024data} & Website, Mobile, Desktop \\ 
        \hline
        FlowGPT & FlowGPT & \url{https://flowgpt.com} & $\geq 4,000,000$ monthly active users \cite{semrush2024data} & Website, Mobile \\ 
        \hline
        Poe & Quora & \url{https://poe.com} & $\geq 27,000,000$ monthly visits \cite{similarweb_poe} & Website, Mobile, Desktop \\ 
        \hline
        Coze & ByteDance & \url{https://www.coze.com} & $\geq 3,000,000$ monthly visits \cite{similarweb_coze} & Website \\ 
        \hline
        Cici & ByteDance & \url{https://www.ciciai.com} & $\geq 2,000,000$ monthly visits \cite{similarweb_ciciai} & Website, Mobile, Browser Extension, Desktop \\ 
        \hline
        Doubao & ByteDance & \url{https://www.doubao.com}  & $\geq 27,450,000$ monthly downloads \cite{bytedance_capcut} & Mobile \\ 
        \hline
        HuggingChat & Hugging Face & \url{https://huggingface.co/chat/assistants} & $\geq 18,000,000$ monthly visits \cite{similarweb_huggingface} & Website \\ 
        \hline
        ChatGLM	& Zhipu AI	&	\url{https://chatglm.cn/main/toolsCenter}	& $\geq 3,000,000$ monthly visits \cite{similarweb_chatglm}	& Website, Mobile	\\ 
        \hline
         ERNIE Bot & Baidu & \url{https://yiyan.baidu.com/agent-square} & $\geq 18,900,000$ monthly visits \cite{similarweb_yiyan}	& Website, Mobile \\ 
         \hline
         Character.AI & Character Technologies & \url{https://character.ai} & $\geq 200,000,000$ monthly visits \cite{similarweb_characterai} & Website \\ 
        \hline
        JanitorAI 	& JanitorAI  & \url{https://janitorai.com} 	& $\geq 45,800,000$ monthly visits \cite{similarweb_janitor}	& Website \\ 
        \hline
        Talkie 	& Minimax & \url{https://talkie-ai.com} & $\geq 4,400,000$ monthly visits \cite{similarweb_talkie}	& Website, Mobile \\ 
        \hline
        Joyland & Westlake Xinchen & \url{http://joyland.ai} & $\geq 3,200,000$ monthly visits \cite{similarweb_joyland}& Website, Mobile \\ 
        \hline
        Chub Venus AI 	& Chub AI	& \url{https://www.chub.ai} 	& $\geq 2,600,000$ monthly visits \cite{similarweb_chub}	& Website, Mobile 	\\ 
        \hline
        Crushon.AI 	& Crushon AI 	& \url{https://crushon.ai} 	& $\geq 11,900,000$ monthly visits \cite{similarweb_crushon}	& Website 	\\ 
    \bottomrule[1.2pt]
    \end{tabular}
    }
    \label{tab:ai_platforms_comparison}
\end{table}

\subsection{Motivations}

As illustrated in \autoref{tab:ai_platforms_comparison}, the rapidly evolving LLM app store ecosystem is experiencing unprecedented growth and diversification. This remarkable expansion is exemplified by the emergence of several innovative key players and the achievement of significant technological and financial milestones within the field.

FlowGPT~\cite{flowgpt} stands as a prime example of the vast untapped potential within LLM app stores, boasting an impressive user base of over 4 million monthly active users. Moreover, it has recently secured a significant milestone by successfully completing a \$10 million Pre-A funding round~\cite{pr2024flowgpt}, underscoring its growing influence and burgeoning success in the competitive sector. Bytedance's Coze~\cite{coze} has also achieved impressive traction in the market, with monthly active users exceeding 3 million~\cite{similarweb_coze}, demonstrating the platform's widespread appeal and user engagement.
Additionally, OpenAI's GPT Store~\cite{gptstore} is leading this technological evolution by hosting an extensive collection of over 3 million diverse and innovative apps~\cite{openai2024gptstore}. In the burgeoning and highly competitive third-party LLM app store arena, as of April 1, 2024, the landscape is diverse and expansive: GPTs App~\cite{gptsappio} dominates with 801,185 apps, and GPTs Hunter~\cite{gptshunter} is not far behind, offering a substantial repository of 519,000 apps. Meanwhile, GPTStore.AI~\cite{gptstoreai} provides a solid selection of 179,895 apps, and GPTs Works~\cite{gptsworks} contributes 103,739 apps, with each platform adding unique value, perspectives, and innovative solutions to the rapidly expanding LLM app ecosystem.
Quora's Poe~\cite{poeexplore}, Hugging Face's HuggingChat~\cite{huggingface_huggingchat}, and Baidu's ERNIE Bot~\cite{baidu_erniebot} have all established themselves as key players in the LLM app store market, attracting significant monthly visits of over 27 million, 18 million, and 18.9 million, respectively, reflecting their strong user engagement and innovative offerings.
Platforms like JanitorAI~\cite{janitor_janitorai}, Talkie~\cite{minimax_talkie}, Joyland~\cite{westlake_joyland}, Chub Venus AI~\cite{chub_chubai}, Crushon.AI~\cite{crushon_crushonai}, Poly.AI~\cite{polyai}, Ohai~\cite{ohai}, and Mona Land~\cite{monaland} are also notable players in the LLM app store market, primarily focusing on providing emotional companionship and interactive experiences tailored to adult users.

\textbf{The rapid expansion of LLM app stores parallels the earlier trajectories observed in traditional mobile app stores}~\cite{harman2012app}, where the proliferation of apps necessitated advanced analytical approaches to ensure quality, security, and relevance. Just as mobile app store analysis has become indispensable in optimizing user experience, app performance, and market dynamics~\cite{Zhu_2024}, a similar emphasis on LLM app store analysis is crucial~\cite{zhang2024look}. This emerging domain presents unique challenges and opportunities, from ensuring the ethical deployment of LLM technologies to navigating the complex dynamics of user engagement and content moderation. \textbf{Given the potential impact and rapid growth of LLM app stores, comprehensive research in this area is of paramount importance.}

\textbf{\underline{Current State of Research.}} Recent studies have begun to address this need by initiating preliminary investigations into LLM app stores, particularly the GPT Store. For example, researchers have examined various aspects, such as the landscape of the LLM app, security threats, and datasets~\cite{zhang2024first,su2024gpt,hou2024gptzoo}. Some studies have explored the ecosystem of GPT Store, emphasizing the need to consider their societal impact, potential biases, misinformation issues, and privacy concerns~\cite{zhao2024gpts,tao2023opening,antebi2024gpt}. To address security risks, Hou~\ea~\cite{hou2024security} proposed a framework for identifying potential threats in LLM apps, such as misleading descriptions, sensitive information collection, and the generation of harmful content. Additionally, Lin~\ea~\cite{lin2024malla} investigated real-world malicious services that have integrated LLMs, underscoring the critical need to confront the cybersecurity challenges posed by the misuse of these powerful models.

While these initial explorations have provided valuable insights, \textbf{the academic landscape in this area remains largely underexplored, presenting an extensive frontier teeming with opportunities for inquiry}. The burgeoning LLM app ecosystem offers a fertile ground for in-depth research that goes beyond these preliminary studies. Comprehensive investigation of LLM app stores remains pivotal for gaining deeper insights into the dynamics of LLM apps in real-world scenarios, encompassing user engagement, market dynamics, and technological trends. Such examination can highlight best practices, pinpoint prevailing challenges, and spotlight areas ripe for enhancement, offering a wealth of research opportunities. Furthermore, delving into LLM app stores can illuminate the broader societal impacts of LLM-driven apps. As these apps gain ubiquity, it becomes imperative to scrutinize their utility, the nature of the content they deliver, and their influence on user choices and behaviors.

\begin{figure*}[!h]
    \centering
    \includegraphics[width=\linewidth]{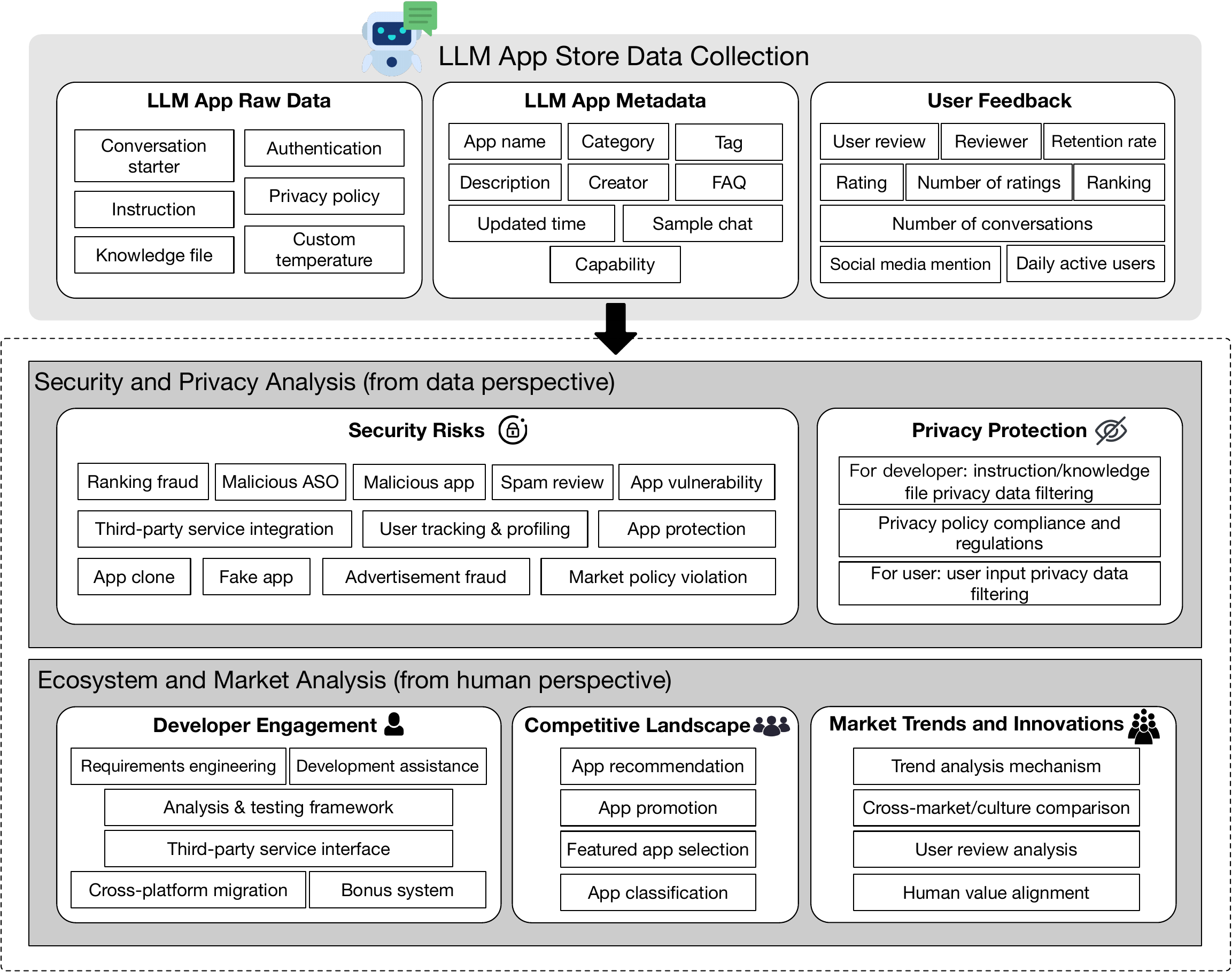}
	\caption{LLM app store mining and analysis roadmap.}
    \label{fig:roadmap}
\end{figure*}

\subsection{Overview}

This paper aims to provide a forward-looking analysis of LLM app stores, focusing on key aspects that shape the user experience, developer strategies, and the dynamics of the ecosystem. Through an exploration of LLM app data, security, privacy, and market dynamics, we aim to uncover trends, pinpoint challenges, and highlight opportunities that could inform future research directions. Rather than proposing a specific framework or solution, this paper serves as a visionary document, highlighting the importance of collaboration and shared responsibility among stakeholders in addressing the challenges and leveraging the opportunities presented by LLM app stores. We believe that by providing insights and recommendations based on a comprehensive analysis of the current landscape, this paper can contribute to the development of a thriving, user-centric, and responsible LLM app ecosystem.

In the following sections, we present the roadmap for mining and analyzing LLM app stores, comprising three key stages as shown in \autoref{fig:roadmap}. The \textit{data collection} stage (\S\ref{sec:data}) involves gathering and preprocessing LLM app raw data, metadata, and user feedback from LLM app stores. The \textit{security and privacy analysis} stage (\S\ref{sec:security}) focuses on identifying potential risks and regulatory compliance issues. The \textit{ecosystem and market analysis} stage (\S\ref{sec:ecosystem}) leverages the collected data to gain insights into developer engagement, market trends, and strategic decision-making within the LLM app ecosystem.
\S\ref{sec:discussion} discusses the implications of our analysis, challenges faced by the ecosystem, and recommendations for stakeholders. Finally, \S\ref{sec:conclusion} concludes the whole paper.

\section{Data collection and preprocessing}
\label{sec:data}
 
To conduct a comprehensive analysis of LLM app stores, researchers must identify and collect key data types, followed by meticulous preprocessing. This section outlines the essential data categories crucial for understanding the LLM app ecosystem, including \textit{LLM app raw data}, \textit{LLM app metadata}, and \textit{user feedback}, as illustrated in \autoref{fig:roadmap}. These data categories provide a multifaceted view of the LLM app landscape, offering insights into app functionality, user engagement, and market trends. The following subsections will delve into each data category, exploring their unique characteristics and analytical potential. Subsequently, we will discuss the critical importance of data preprocessing in ensuring data quality and preparing the dataset for robust analysis. 

\subsection{LLM app raw data}

LLM app raw data encompasses various components that define the behavior and capabilities of the LLM apps. \textbf{Instructions} play a vital role in specifying the desired functionality and behavior of the app, outlining actions to perform and those to avoid. \textbf{Knowledge files} provide custom information that the LLM app can access to inform its responses, retrieving relevant sections based on user input. These files may be viewable by other users through LLM app responses or citations, enhancing transparency and trust. 

When LLM apps require integration with \textbf{\textit{third-party services}}, \textbf{authentication mechanisms} such as API keys or OAuth protocols~\cite{leiba2012oauth} are essential to ensure secure access. Additionally, LLM apps must adhere to the \textbf{privacy policies} of any integrated third-party platforms to maintain user confidentiality. 

\textbf{Conversation starters} are designed to guide new users in asking better questions, providing a smooth onboarding experience. Lastly, \textbf{custom temperature} settings allow for controlling the creativity of the LLM app's responses, balancing variation and predictability to suit different use cases.

\subsection{LLM app metadata}

LLM app metadata plays a crucial role in helping users navigate the LLM app store, providing essential information about each app to facilitate discovery, understanding, and comparison. The \textbf{app name} and \textbf{creator} are fundamental pieces of metadata, allowing users to identify and attribute each app to its respective developer. A detailed \textbf{description} of the app's purpose and features is essential for users to grasp the app's intended use case and capabilities quickly. 
\textbf{Capabilities} provide users with a clear understanding of the app's functionalities. They can include a wide range of features, such as web browsing, which enables the app to access and retrieve information from the internet; image generation, allowing users to create visual content through the app; and code interpretation, enabling the app to understand and execute programming languages. Other potential capabilities include speech generation, video generation, etc. 

\textbf{Categories} organize LLM apps according to their primary function or domain, such as productivity, entertainment, or education. \textbf{Tags} offer more specific details about the app's features, use cases, or compatibility. For instance, tags might indicate whether an app is beginner-friendly, supports multiple languages, or specializes in code generation. The \textbf{updated time} keeps users informed about the app's recency, ensuring they have access to the latest features and content.
\textbf{Sample chats} demonstrate the app's conversational capabilities, response quality, and potential applications, providing users with a realistic preview of the experience. \textbf{Frequently Asked Questions (FAQs)} serve as a crucial component that methodically addresses common user inquiries. This segment offers concise answers to typical questions about the app's functionality, limitations, and recommended practices.

\subsection{User feedback}

User feedback is a valuable source of data for assessing the performance and popularity of LLM apps. One of the key metrics is the \textbf{number of conversations}, which indicates the level of user engagement with the app. A high number of conversations suggests that users find the app valuable and engaging, regularly interacting with it to fulfill their needs. The \textbf{retention rate} measures the percentage of users who continue to use the app over a specific period. \textbf{Daily active users (DAU)} provide a snapshot of the app's active user base, representing the number of unique users who engage with the app daily. Tracking DAU over time offers insights into the app's ongoing appeal and growth trajectory. 

\textbf{Ratings} and the \textbf{number of ratings} offer a quantitative measure of user satisfaction, allowing users to express their opinions on a standardized scale. A high average rating and a large number of ratings signify that users generally have a positive experience with the app and are willing to share their feedback. \textbf{Rankings} provide a comparative measure of an app's performance against other similar apps within the store. 

\textbf{User reviews} offer qualitative feedback, allowing users to share detailed opinions, experiences, and suggestions. Positive reviews highlight an app's strengths and the value it provides to users, while negative reviews can reveal weaknesses, bugs, or areas for improvement. Analyzing user reviews can help developers prioritize updates, fix issues, and enhance features based on user preferences. Information about \textbf{reviewers}, such as their user profile or history with the app, can provide additional context and credibility to their feedback. \textbf{Social media mentions} capture an LLM app's broader impact and popularity beyond the confines of the LLM app store. Users may share their experiences, recommend the app to others, or engage in discussions related to the app on various social media platforms. 

\subsection{Data preprocessing}

Once the data is collected, it must undergo a rigorous preprocessing phase to ensure its quality, security, and compliance with privacy regulations. Preprocessing steps should be applied to ensure data quality and consistency~\cite{alasadi2017review,garcia2015data,mishra2020new}. This involves removing duplicate entries, handling missing values, and normalizing text data. Text preprocessing techniques such as tokenization, lowercasing, and removing stop words and punctuation should be employed. Data cleaning steps, such as removing irrelevant or spam reviews and filtering out apps with insufficient information or user engagement, are also necessary. This preprocessing phase is crucial for obtaining high-quality, reliable data for analysis. Furthermore, this phase may involve filtering out sensitive or personal information, removing malicious content, and ensuring adherence to established policies and guidelines. Data normalization and formatting procedures are also applied to facilitate efficient storage, retrieval, and analysis of the collected information.
By following this comprehensive data collection and preprocessing approach, researchers can gain a holistic understanding of the LLM app ecosystem, enabling them to conduct in-depth analyses, identify trends and patterns, and ultimately contribute to the advancement and growth of this rapidly evolving field.

\section{Security and privacy analysis}
\label{sec:security}

As shown in \autoref{fig:roadmap}, in the evolving landscape of LLM app stores, security and privacy emerge as paramount concerns, necessitating a comprehensive and multifaceted analysis to ensure the integrity and trustworthiness of the ecosystem.

\subsection{Security risks}

\noindent\textit{\textbf{\underline{LLM app raw data-related risks.}}}
\textbf{App cloning}, where someone unauthorized copies a legitimate app, infringes on intellectual property rights and potentially introduces security threats or subpar user experiences. In the mobile app ecosystem, app cloning has been a persistent issue, and app stores have employed techniques like code signing~\cite{cooper2018security,cho2012double} and similarity analysis~\cite{al2019empirical,gonzalez2015droidkin} to detect and prevent cloned apps. However, in the context of LLM apps, cloning challenges differ due to the reliance on proprietary base LLMs and unique prompt engineering strategies. This necessitates the development of specialized detection mechanisms that account for these distinctive features, such as monitoring for duplicated prompt patterns or misuse of LLMs. For LLM app stores, researchers should explore similarity analysis techniques tailored to LLM apps, to combat app cloning effectively. 

\textbf{App vulnerabilities} refer to security weaknesses within LLM apps that attackers can exploit, potentially leading to data breaches or unauthorized activities. For example, these vulnerabilities could arise from inadequate input validation, allowing attackers to perform injection attacks with crafted inputs to elicit unintended responses. Unlike traditional mobile apps, LLM apps are particularly susceptible to prompt injection attacks and adversarial inputs that exploit the language understanding capabilities of LLMs, requiring novel defense mechanisms specifically designed to address these challenges. This can manipulate LLMs into generating sensitive information, violating content policies, or executing unauthorized actions, compromising app integrity and user data. Additionally, insufficient input checks may render apps vulnerable to jailbreaking, enabling LLMs to output content or perform tasks against terms of service or regulations, raising legal and ethical concerns.

These security flaws are also frequently the result of substandard development practices, such as insecure data storage, where sensitive information is poorly protected, making it accessible to unauthorized parties. Weak encryption methods or a lack of robust database security can further exacerbate these issues. Moreover, inadequate authentication mechanisms, including predictable passwords or the absence of multi-factor authentication, can simplify unauthorized access to app functionalities and data. 
App vulnerabilities are not uncommon in the mobile app ecosystem~\cite{mutchler2015large,watanabe2017understanding}, with a wealth of established detection techniques available~\cite{abdullah2021android,chen2022ausera,shezan2017vulnerability}. Accordingly, one of the future research directions should be developing tailored solutions to identify and mitigate vulnerabilities specific to LLM apps. This may involve techniques for securing input validation, preventing jailbreaking, enforcing robust authentication, and ensuring secure data storage and transmission within the LLM app ecosystem.

\textbf{Malicious apps} pose a substantial risk in LLM app stores. The malice can manifest in several ways. For example, developers may create LLM apps using instructions or knowledge files containing malicious content, resulting in the app's knowledge base being tainted with harmful information. Moreover, LLM apps themselves may output content lacking proper constraints, including pornographic or gambling-related information, or even links directing users to malicious websites. Another example is low description-to-behavior fidelity, that is when the actual performance or actions of an app diverge significantly from its documented descriptions or expected behaviors. These phenomena are also prevalent in the mobile app ecosystem. Various tools and techniques have been developed to detect and mitigate malicious mobile apps, such as static and dynamic analysis~\cite{martinelli2016find,quan2014detection,seo2014detecting}, machine learning-based malware detection~\cite{li2018significant,soni2020malicious,wei2017machine}, and app vetting processes~\cite{backes2014taking,zhang2013vetting}. Given that LLM apps can dynamically generate content, malicious actors might exploit this capability to produce harmful outputs that are harder to detect using traditional static analysis, emphasizing the need for advanced real-time monitoring and content validation methods unique to LLM environments. The unique challenges posed by malicious LLM apps necessitate the development of tailored detection and mitigation strategies. Researchers should focus on developing novel techniques specifically designed to identify and address the distinct threats posed by malicious LLM apps, ensuring a safe and trustworthy ecosystem.                           

\textbf{Third-party service integration} is another area of concern, as integrating external services or APIs into an app can introduce vulnerabilities or data privacy issues. For example, if the third-party service provider experiences a data breach or has weak security measures, it could compromise the security and privacy of the LLM app and its users. In the mobile app domain, various mature methods have been proposed to address similar issues, including extensive research on third-party library analysis~\cite{feal2021don,utz2022privacy,he2020diversified,kawabata2013sanadbox}.
To effectively mitigate the risks associated with third-party service integration in LLM apps, developers should adhere to the principle of least privilege, granting only the minimum necessary permissions and access required for the service to function within the app. Robust authentication and authorization mechanisms should be implemented to ensure that only authorized users and processes can interact with the integrated services. Furthermore, encrypting sensitive data both in transit and at rest is crucial when exchanging information with third-party services to protect the confidentiality and integrity of the data. Regular monitoring and auditing of third-party services should also be conducted to detect any suspicious activities or changes in their security posture. Additionally, the interactive and adaptive nature of LLM apps may inadvertently expose sensitive user interactions to third-party services, necessitating more stringent data handling policies and privacy-preserving integrations specifically designed for the LLM context.

\textbf{User tracking and profiling} without proper consent is another risk, where excessive tracking of user data, behavior, or activities occurs, often for targeted advertising or analyzing user preferences. 
This can manifest in various harmful ways, such as identity theft, personalized phishing attacks, or unwanted exposure to tailored yet intrusive advertising~\cite{li2023you,wei2012profiledroid,yalcin2017extracting}. Unlike conventional apps, LLM apps may infer additional personal information from seemingly innocuous user inputs due to their advanced language understanding, amplifying privacy concerns and highlighting the need for specialized safeguards against unintended data collection and profiling. Moreover, the accumulation and analysis of such data could result in biased or discriminatory outcomes, where decisions made by these LLM apps might favor or disfavor individuals based on their profiled characteristics. This not only undermines user trust but also raises ethical concerns about the fairness and transparency of LLM-powered apps. To mitigate risks associated with user tracking and profiling, LLM app stores should enforce strict privacy policies, obtain explicit user consent, and employ privacy-preserving techniques like differential privacy and data anonymization. Strategies such as regular audits should be adopted to ensure fairness, accountability, and transparency in LLM app decision-making.

Similar to mobile app protection techniques that often involve obfuscation~\cite{continella2017obfuscation,wang2018software}, encryption~\cite{silva2013data,tysowski2013hybrid}, and packing~\cite{chen2018mobile}, LLM apps may employ comparable \textbf{app protection} techniques to safeguard their models and data. For example, a potential risk arises from the reliance on third-party frameworks for app protection. To safeguard against model stealing~\cite{tramer2016stealing,orekondy2019knockoff} and unauthorized model reuse~\cite{ji2018model}, developers might obfuscate their LLM apps to protect the model itself. However, this obfuscation process could introduce new security risks. It might inadvertently obscure crucial monitoring and debugging features, making it harder to identify and respond to genuine security threats. Additionally, the complexity added by obfuscation could lead to performance degradation, not only affecting the user experience but also potentially introducing vulnerabilities that attackers could exploit.

\textbf{Advertisement fraud} can occur during the user's interaction with the LLM app, involving deceptive or misleading ad practices, such as hidden payments, unauthorized data collection, or intrusive ad experiences. Mobile app stores have employed ad network monitoring~\cite{haddadi2010mobiad}, real-time ad analysis~\cite{shao2018understanding}, and user feedback analysis~\cite{gui2015truth} to combat advertisement fraud. However, LLM apps can dynamically generate content, including advertisements, which may be manipulated to display fraudulent or malicious ads that are harder to detect using conventional monitoring techniques. This necessitates advanced AI-driven solutions that can analyze and interpret generated content in real-time, specifically tailored to the generative capabilities of LLMs. For LLM app stores, researchers should explore adapting these techniques and developing new methods tailored to the unique challenges of LLM apps, ensuring a transparent and trustworthy advertising ecosystem.

\textbf{Market policy violations}, where LLM apps breach the LLM app store's terms of service, content policies, or other regulations governing app publication and monetization, can undermine the LLM app store's integrity and user trust. Mobile app stores have implemented automated policy compliance checks~\cite{liu2021have,zimmeck2016automated} and app vetting processes~\cite{zhang2013vetting} to enforce market regulations. Due to the generative and adaptive capabilities of LLMs, policy violations may occur dynamically during user interactions, requiring continuous monitoring and compliance enforcement mechanisms that are specifically designed for the evolving outputs of LLM apps. In the context of LLM app stores, researchers should focus on developing automated policy compliance checks tailored to the unique characteristics and challenges of LLM apps, ensuring a secure and trustworthy LLM app ecosystem.

\noindent\textit{\textbf{\underline{LLM app metadata-related risks.}}}
\textbf{Fake apps}, designed to impersonate legitimate LLM apps and deceive users or steal sensitive information, pose a significant risk to users. Mobile app stores have implemented app vetting processes~\cite{zhang2013vetting} and leveraged techniques like app analysis~\cite{rani2023fake} and user feedback~\cite{martens2019towards} monitoring to identify fake apps. Given the sophisticated language capabilities of LLMs, fake apps can produce highly convincing content and interactions, making them more deceptive than traditional fake apps and requiring more nuanced detection strategies that consider the quality and context of generated outputs. In the context of LLM app stores, researchers should investigate developing advanced natural language processing and multimedia analysis methods to aid in the detection of fake LLM apps, ensuring user safety and trust.

\noindent\textit{\textbf{\underline{User feedback-related risks.}}}
In the context of LLM app stores, security risks should be carefully considered and addressed. One significant risk is \textbf{ranking fraud}, where attackers attempt to manipulate the LLM app store rankings through illegal methods, such as using bot programs to generate fake ratings, downloads, or reviews, or engaging in keyword stuffing. Similar to the mobile app market, where researchers have proposed systems to detect ranking fraud by analyzing leading sessions, rating patterns, and review behaviors~\cite{zhu2013ranking,zhu2014discovery}, app stores may need to employ advanced techniques to identify and mitigate fraudulent activities aimed at artificially inflating app rankings and popularity. Moreover, attackers might leverage LLMs themselves to generate human-like fake reviews and feedback at scale, complicating detection efforts and necessitating the development of sophisticated algorithms capable of distinguishing between genuine and AI-generated content.

Another concern is \textbf{malicious ASO} (i.e., App Store Optimization), where attackers exploit irregular methods to falsify user feedback, such as user engagement metrics or app ratings, to artificially boost an app's search result rankings and discoverability, ultimately gaining higher exposure and usage when users search for related keywords. This issue is analogous to the collusive promotion groups in the mobile app ecosystem, where developers pay service providers to organize groups of attackers to post fraudulent reviews, inflate download numbers, or manipulate app ratings in an attempt to boost their app's ranking and visibility~\cite{chen2017toward,padilla2019importance,wang2019understanding}, which can ultimately undermine the integrity of the app store's ecosystem if not addressed.

\textbf{Spam reviews} in LLM app stores can also contain malicious content or involve large-scale fake reviews manipulated by bots or manual efforts, intending to inflate the app's reputation artificially. This issue is well-documented in the mobile app industry, where spam reviews and review fraud have been a persistent challenge. Various detection methods have been established in the mobile app domain~\cite{genc2019detection,seneviratne2015early}. Similarly, LLM app stores must adopt and refine such techniques to preserve the authenticity and reliability of their review mechanisms. Considering that LLMs can generate high-quality text, spam reviews may become more sophisticated to distinguish from genuine user feedback, requiring enhanced detection methods that analyze linguistic patterns and contextual relevance specific to LLM-generated content.

\subsection{Privacy protection}

Protecting privacy is a critical aspect of LLM app stores. \textbf{For developers}, it is essential to filter out any privacy data that may be included in the instructions or knowledge base files provided to the LLM app. This includes not only personally identifiable information (PII)~\cite{schwartz2011pii} such as addresses, contact details, and other sensitive data that could compromise individual user privacy, but also extends to sensitive information related to businesses, governmental bodies, and other entities. Protecting this wider range of data ensures the privacy and security of related stakeholders, safeguarding against potential misuse, data breaches, or other forms of exploitation that could have far-reaching consequences. This is similar to the principles and practices adopted in the mobile app industry, where developers are required to implement appropriate data protection measures to safeguard user privacy and comply with relevant regulations, such as the General Data Protection Regulation (GDPR)~\cite{voigt2017eu}. 

Furthermore, developers must also comply with the LLM app store's \textbf{privacy policies} and relevant legal regulations when collecting user information for personalized fine-tuning or optimization of the app. This involves clearly informing users about the purpose, scope, and manner of data collection and usage, and obtaining user consent. The LLM app store should review the app to ensure compliance with these requirements. Again, this aligns with the standard practices in mobile app stores, where apps are vetted for their data collection and privacy practices, and users are provided with clear information about how their data is being used~\cite{almuhimedi2015your,shklovski2014leakiness}. Given that LLM apps may process large volumes of unstructured user-generated text, which can inadvertently contain sensitive information, developers must implement advanced consent mechanisms and data handling policies that address the specific risks associated with natural language data processing, ensuring that users are adequately informed and protected.

\textbf{From the user's perspective}, privacy data filtering is crucial when interacting with LLM apps. Users' input may contain private information, and the app should have filtering mechanisms in place to identify and remove this sensitive data, preventing it from being leaked to developers or stored in the knowledge base. This is analogous to the privacy protection measures implemented in mobile apps, where user inputs and data are often processed locally on the device or through secure channels to protect user privacy~\cite{liu2014reconciling,nema2022analyzing}. However, the conversational and predictive capabilities of LLMs mean that they might generate outputs based on sensitive user data, even when not explicitly provided in a single interaction, requiring innovative privacy safeguards that prevent the unintentional disclosure of personal information through model predictions. Additionally, users can expect LLM app stores to provide transparent information about the privacy practices of listed apps, similar to how mobile app stores provide privacy labels and summaries to help users make informed decisions~\cite{apple2024privacy}.

\section{Ecosystem and market analysis}
\label{sec:ecosystem}

In the dynamic ecosystem of LLM app stores, the interplay between developer engagement, competitive landscape, and market trends drives innovation and growth. As displayed in \autoref{fig:roadmap}, developer support mechanisms, strategies for navigating competitive pressures, and responsiveness to evolving market dynamics are crucial for cultivating a vibrant, sustainable marketplace that caters to diverse user needs and preferences while fostering technological advancements.

\subsection{Developer engagement}

Enhancing support for LLM app developers is essential to fostering a thriving ecosystem. Implementing effective \textbf{requirements engineering} processes and tools can help developers gain clarity on app specifications and functionalities. Although mobile app development benefits from established practices like user story mapping~\cite{cohn2004user} and wireframing~\cite{brown2010communicating}, the LLM app ecosystem should develop specialized tools that cater to the unique needs of conversational AI, such as dialogue flow designers, intent mappers, and entity recognizers.

Providing comprehensive \textbf{development assistance}, including documentation, examples, and best practices, can lower entry barriers and guide developers in creating high-quality LLM apps. Drawing inspiration from the mobile app domain's extensive resources, such as Apple's Human Interface Guidelines~\cite{apple2021human} and Google's Material Design~\cite{google2021material}, the LLM app ecosystem should create similar guides tailored to conversational AI, covering topics like prompt engineering~\cite{arvidsson2023prompt}, context management~\cite{ge2023context}, and multi-turn dialogue handling~\cite{wu2023autogen}.

Offering robust \textbf{analysis and testing tools or frameworks} can assist developers in evaluating app performance, identifying vulnerabilities, and optimizing the user experience, ensuring high-quality output. Mobile app development has benefited from tools like Appium~\cite{verma2017mobile} and Espresso~\cite{google2021espresso}, which have revolutionized automated testing, enabling developers to catch bugs early and ensure app stability. Similarly, the LLM app ecosystem needs to invest in developing comprehensive testing frameworks that can effectively simulate user interactions and detect potential biases or inconsistencies in the generated output, thus improving overall reliability and user satisfaction.

Standardized \textbf{third-party service interfaces} can simplify the integration process for developers. The LLM app store can provide a list of certified service providers or establish partnerships with leading companies in LLMs and knowledge bases, similar to how mobile app stores have streamlined integration with payment gateways~\cite{isaac2012anonymous,yang2019security} and analytics~\cite{han2016mobile,minelli2013software} providers.

\textbf{Cross-platform migration tools} and support can help developers deploy LLM apps across multiple platforms. In the mobile app development domain, frameworks like React Native~\cite{eisenman2015learning} and Flutter~\cite{kuzmin2020experience,wu2018react} have greatly simplified the process of building cross-platform apps. The LLM app ecosystem could explore similar solutions that allow developers to write once and deploy across various conversational AI platforms.

Implementing a comprehensive \textbf{bonus system} that rewards developers based on app quality and user feedback can incentivize continuous optimization. Similar to Apple's app store small business program~\cite{apple2021smallbusiness} offering financial incentives and recognition for high-performing developers, the LLM app ecosystem should consider initiatives that encourage innovation and user satisfaction.

\subsection{Competitive landscape}

Leveraging user preferences, history, and ratings, LLM app stores can develop sophisticated \textbf{recommendation algorithms} to suggest potentially interesting LLM apps to users. This not only improves app discoverability but also increases user satisfaction and engagement. Mobile app stores have successfully implemented such recommendation systems, with examples like Apple's app store featuring ``Apps You Might Like'' and Google Play's ``Recommended for You'' sections~\cite{tu2019personalized, hwang2019mobile}.
However, in the context of LLM app stores, academic research on tailoring recommendation algorithms to this novel domain remains largely unexplored.

To help developers effectively promote their LLM apps, LLM app stores can offer a range of \textbf{promotion tools and channels}. Mobile app stores usually use ads, featured spots, and events to promote mobile apps, leveraging influencer partnerships for added trust and visibility~\cite{lee2018mobile,rahman2019art}. Similarly, LLM app stores could also include advertising placement options, such as sponsored search results or featured app listings, allowing developers to increase their app's visibility to potential users. Additionally, LLM app stores can provide promotional opportunities through curated collections, themed showcases, or developer spotlights, highlighting noteworthy LLM apps and their creators.

\textbf{Curating and featuring} top LLM apps is a strategic move by LLM app stores to influence the app market landscape~\cite{gptshunter,gptstore,gptstoreai}. LLM app stores could establish benchmarks and inspire developers to aim high by spotlighting apps that excel in innovation and quality. This not only guides users to superior apps but also rewards developers for outstanding user experiences. 

LLM app stores are encouraged to deploy a coherent \textbf{app classification} system to improve LLM app discoverability. Sorting LLM apps by their functionality, usage scenarios, industries, or target audiences simplifies the search process for users. This not only elevates the user experience but also supports developers in strategically showcasing their apps. For instance, renowned stores like Google Play~\cite{fu2013people,googleplay}, Apple App Store~\cite{appleapp,jia2020smartphone}, and Blackberry World App Store~\cite{harman2012app} typically employ a consolidated category structure. 

\subsection{Market trends and innovations}

Establishing a \textbf{trend analysis} mechanism is crucial for LLM app stores to uncover and predict LLM app market trends by mining user behavior data, download volumes, and reviews. This helps app stores and developers formulate future strategies, such as identifying increasingly popular features or scenarios. LLM app stores can draw inspiration from the well-established practices in traditional mobile app stores, which have successfully employed trend analysis techniques to identify emerging app trends and user preferences~\cite{fu2013people,liu2015personalized,milward2016user}. For instance, analyzing in-app user behavior patterns and feature usage can provide valuable insights into user preferences and emerging trends for LLM apps.

Analyzing and comparing consumer preferences for LLM apps \textbf{across different regional markets and cultural backgrounds} can reveal market differences, enabling LLM app stores to adjust product strategies and operational approaches accordingly. Cross-cultural research on mobile app adoption has highlighted the importance of tailoring app interfaces, content, and functionality to cater to diverse cultural norms and expectations, which can significantly impact user engagement and retention~\cite{light2018walkthrough,sultan2009factors}. LLM app stores can leverage these learnings and adapt their offerings, marketing strategies, and localization efforts to better resonate with users from various cultural backgrounds.

\textbf{User review analysis} is a vital channel for LLM app stores to understand genuine user feedback and identify areas for improvement. By applying natural language processing and sentiment analysis to a vast number of user reviews, stores can gain insights into user pain points, app deficiencies, bugs, user expectations, suggestions, and overall acceptance and trust levels for LLM apps. Just as in the mobile app domain, where user reviews have been extensively leveraged to improve app quality and user experience~\cite{guzman2014users}, LLM app stores can benefit from similar techniques to mine valuable feedback from user reviews.

As artificial intelligence apps, the development of LLM apps must \textbf{adhere to human ethical values} and maintain a high degree of alignment with humanistic ideals, which is essential for gaining public trust and recognition. App stores should establish review standards that prohibit the listing of LLM apps containing content that violates social morality or harms public interests. The design of LLM apps should embody human-centric values, such as respect for privacy, explainability, and controllability. The functions, algorithms, knowledge bases, and other aspects of LLM apps must align with human interests and avoid producing harmful effects. Similar to the guidelines and best practices established in the mobile app industry for protecting user privacy, ensuring data security, and promoting ethical app development~\cite{balebako2014privacy,martinez2015privacy}, LLM app stores can adopt and adapt these principles to address the unique challenges and risks associated with AI-powered apps.

\section{Discussion}
\label{sec:discussion}

This section analyzes the implications of the proposed roadmap for LLM app store development and regulation. It also discusses key challenges and provides recommendations for LLM app store stakeholders.

\subsection{Implications}

The burgeoning LLM app store ecosystem presents a unique blend of opportunities and challenges that have far-reaching implications for developers, users, regulators, and the broader AI community. For developers, the potential to create innovative, AI-driven apps is immense, opening avenues for novel applications in education, healthcare, finance, and entertainment. This potential comes with the responsibility to ensure that apps are secure, privacy-compliant, and ethically aligned. Detailed analysis of LLM app raw data, metadata, and user feedback is crucial for developers to understand user needs and preferences, enabling them to design more engaging and useful apps that adhere to high standards of quality and reliability.
For users, the proliferation of LLM apps enhances access to advanced AI capabilities, democratizing technology and fostering greater engagement with digital platforms. However, this increased accessibility also raises concerns about the potential for misuse, such as the spread of misinformation or the erosion of privacy. Users must navigate the benefits and risks associated with LLM apps, underscoring the need for transparency and education to make informed decisions.
For regulators and LLM app store managers, the rapid evolution of LLM app stores necessitates a proactive approach to governance. Ensuring a safe, trustworthy, and inclusive platform requires continuous monitoring for security threats such as malicious apps, spam reviews, and ranking fraud. Privacy protection remains paramount, demanding stringent measures to safeguard user data from unauthorized tracking, profiling, and third-party service vulnerabilities. The development of appropriate legal frameworks and industry standards is critical to balance innovation with the protection of user rights.

Moreover, the growth of LLM app stores has significant implications for the broader AI ecosystem. The rich data generated by user interactions provides valuable insights for researchers and can drive advancements in AI technologies. However, this also places a responsibility on all stakeholders to address ethical considerations, such as algorithmic biases and fairness, to ensure that the benefits of LLM apps are equitably distributed across society.

\subsection{Challenges}

The analysis of LLM app stores following the proposed roadmap reveals several challenges that need to be addressed to ensure the sustainable growth and responsible development of this ecosystem.

\noindent\textbf{Data privacy and security.} The integration of third-party services and the collection of user data by LLM apps raise significant privacy and security concerns. Ensuring compliance with data protection regulations, such as GDPR and CCPA, and implementing robust security measures to prevent data breaches and unauthorized access to user information are critical challenges that require attention from both developers and platform providers. Additionally, the dynamic and adaptive nature of LLMs may inadvertently collect sensitive user information during interactions, necessitating sophisticated mechanisms to anonymize and protect user data in real-time.

\noindent\textbf{Intellectual property protection.} The prevalence of app cloning and the potential for intellectual property infringement within LLM app stores pose significant challenges to developers and platform owners. Detecting and preventing the unauthorized copying or reuse of app code, designs, and features is crucial to maintaining a fair and competitive environment that rewards innovation and original work. Furthermore, the generative capabilities of LLMs can complicate intellectual property issues, as they may produce content resembling existing works, raising questions about authorship and originality.

\noindent\textbf{Ensuring app quality and reliability.} With the rapid growth of LLM app stores, maintaining high standards of app quality and reliability becomes increasingly challenging. Implementing effective app review processes, establishing clear guidelines for developers, and continuously monitoring app performance and user feedback are essential to provide users with a consistent and trustworthy experience. The inherent unpredictability of AI-generated outputs necessitates rigorous testing and validation procedures to prevent unintended behaviors or harmful content.

\noindent\textbf{Addressing algorithmic biases and fairness.} LLM apps rely on complex algorithms and models that may inadvertently perpetuate biases or discriminate against certain user groups. Identifying and mitigating these biases, ensuring fairness in app recommendations and search results, and promoting diversity and inclusivity within the app ecosystem are significant challenges that require ongoing research and collaboration between developers, researchers, and LLM app store managers. The lack of transparency in AI decision-making processes further complicates efforts to detect and correct biases.

\noindent\textbf{Balancing innovation and responsibility.} The rapid advancements in LLM technologies and the increasing capabilities of LLM apps present both significant opportunities for innovation and formidable challenges in terms of responsible development and deployment. Striking the right balance between pushing the boundaries of what is possible and considering the ethical, social, and long-term implications of LLM apps is a critical challenge that requires input from multiple stakeholders, including developers, researchers, policymakers, and users. Policies and practices must encourage innovation while imposing necessary constraints to prevent misuse.

\noindent\textbf{User education and awareness.} As LLM apps become more prevalent and influential in various domains, educating users about their capabilities, limitations, and potential risks becomes increasingly important. Providing clear and accessible information about how LLM apps work, what data they collect, and how users can control their interactions with these apps is a significant challenge that requires collaboration between developers, platform providers, and educational institutions. User trust hinges on transparency and the ability to make informed decisions about app usage.

\noindent\textbf{Regulatory and policy challenges.} The rapid growth and evolving nature of the LLM app ecosystem present challenges for regulatory bodies and policymakers. Developing appropriate legal frameworks, guidelines, and standards that promote innovation while protecting user rights and ensuring accountability is a complex task that requires ongoing dialogue and collaboration between industry stakeholders and policymakers. Issues such as liability for AI-generated content, cross-border data flows, and international cooperation add layers of complexity to the regulatory landscape.

\subsection{Recommendations for LLM app store stakeholders}

Recognizing the need for more concrete and actionable recommendations, we provide the following suggestions tailored to each stakeholder to drive future research and practical implementation in the field.

\noindent\textbf{LLM app store managers.}
For LLM app store managers, implementing concrete measures is crucial to fostering a secure and innovative ecosystem. A primary focus should be on developing a comprehensive submission process that addresses the unique challenges of LLM apps. This process should include a detailed checklist covering critical aspects such as data encryption methods, third-party service security certifications, bias mitigation efforts, and user data handling practices. By requiring developers to provide this information, managers can ensure a thorough evaluation of each app's security and ethical considerations.
To streamline this process, managers should invest in developing automated analysis tools tailored specifically for LLM apps. These tools could scan for common vulnerabilities in LLM integration, data handling, and third-party service usage, providing developers with immediate feedback to address potential issues before submission. Making these tools available to developers pre-submission can significantly improve the overall quality and security of apps in the ecosystem.
Implementing a tiered review system based on an app's potential risk level can further enhance the vetting process. Apps handling sensitive data or utilizing advanced AI features should undergo more rigorous scrutiny, potentially including mandatory communications with developers to discuss security and ethical considerations in depth. This approach ensures that high-risk apps receive the attention they require while maintaining an efficient review process for lower-risk submissions.

Transparency should be a key priority for LLM app store managers. Creating a public dashboard that displays real-time statistics on app review processes, including average review times, common rejection reasons, and trends in security issues, can provide valuable insights to developers and users alike. This transparency not only guides developers in improving their submissions but also builds trust within the ecosystem. To incentivize and recognize excellence in security and ethical practices, managers should consider implementing a certification program for apps that meet stringent standards. This certification, visibly displayed to users, can serve as a mark of trust and quality, encouraging developers to prioritize these crucial aspects in their app development process.

Supporting the developer community is another vital role for LLM app store managers. This support can take the form of open-source tools designed specifically for LLM app development, such as bias detection toolkits and modules to enhance the explainability of LLM decisions to users. By providing these resources, managers can actively contribute to raising the overall standard of apps in their ecosystem.
Engaging with the broader security community is equally important. Establishing a bug bounty program focused on LLM apps can incentivize security researchers to identify vulnerabilities unique to this ecosystem, thereby continuously improving the platform's security posture.

To address the complex ethical challenges posed by LLM apps, managers should consider forming an ethics advisory board comprising experts in AI ethics, security, and law. This board can provide guidance on controversial apps and emerging ethical dilemmas, ensuring that the app store's policies evolve in step with technological advancements. Fostering collaboration between industry and academia is crucial for addressing long-term challenges in the field. Launching research grants focused on LLM app security and ethics can drive innovation and provide valuable insights for future policy and technology development. Leveraging AI technologies to enhance user feedback mechanisms can provide valuable insights into potential security and ethical concerns. Implementing natural language processing systems to analyze user reviews can help identify emerging issues, allowing both developers and the app store team to address concerns proactively.

\noindent\textbf{LLM app developers (creators).}
For developers, implementing specific measures is crucial to ensure their LLM apps are secure, ethical, and user-centric. They should adopt a comprehensive security-first approach, implementing robust authentication and encryption protocols for third-party services and data transmission. Developers must carefully design prompts and system messages to guide the base model's behavior, ensuring that outputs align with the app's intended functionality and ethical standards. To address potential biases, developers should implement comprehensive testing protocols, using diverse input scenarios to identify and mitigate any unintended biases in the app's responses. Creating easily accessible in-app mechanisms for users to report problematic outputs is essential for continuous improvement. Transparency can be enhanced by providing clear documentation on the app's capabilities, limitations, and the specific ways it utilizes the base model. To improve user satisfaction, analytics on user interactions and feedback can be used to refine prompts and app functionality. 

\noindent\textbf{Researchers.}
Researchers should conduct in-depth studies on user behavior, market trends, and security challenges specific to LLM apps, providing valuable insights for all stakeholders. For instance, they could analyze user interaction patterns with different types of LLM apps to identify potential risks and opportunities for improvement. Developing standardized evaluation metrics for LLM app performance, safety, and ethical compliance could help establish industry benchmarks. Researchers should also focus on creating tools and methodologies for detecting and mitigating biases in LLM app outputs, considering the unique challenges posed by LLM apps.

\noindent\textbf{Policymakers.}
Policymakers need to develop adaptive regulatory frameworks that address the distinct characteristics of LLM apps. This includes crafting guidelines for data privacy, user consent, and content moderation that are tailored to the dynamic nature of LLM apps. Collaboration between researchers and policymakers is crucial in developing ethical AI use frameworks that can be practically implemented by developers and enforced by app store managers. These frameworks should address issues such as transparency in AI-generated content, accountability for app outputs, and safeguards against potential misuse. 

\noindent\textbf{Users.}
For users, staying informed about the apps they use is key. This includes reviewing app permissions, understanding data usage policies, and providing constructive feedback to developers. By actively participating in the ecosystem, users can contribute to the improvement of LLM apps and help foster a culture of transparency and accountability.

The landscape of LLM app stores presents a unique confluence of innovation, opportunity, and responsibility. Our analysis underscores the critical need for a coordinated, multi-stakeholder approach to navigate the complexities of this rapidly evolving ecosystem. By implementing the proposed strategies, LLM app store managers can foster a secure and transparent environment, developers can create more responsible and user-centric apps, researchers can drive evidence-based improvements, and policymakers can craft adaptive regulatory frameworks. Users, empowered with knowledge and tools, become active participants in shaping the ecosystem's trajectory.

\section{Conclusion}
\label{sec:conclusion}
This paper provides a forward-looking analysis of LLM app stores, focusing on key aspects such as app data collection, security and privacy analysis, and ecosystem and market analysis. Through this exploration, we underscore the importance of user-centric design, data privacy, intellectual property protection, and collaboration among stakeholders in shaping the future of the LLM app ecosystem.
As the LLM app landscape continues to evolve, ongoing research and collaboration among researchers, developers, LLM app store managers, and policymakers are crucial to address challenges, leverage opportunities, and drive responsible innovation.

\section*{ACKNOWLEDGMENT}
This work was supported in part by the Key R\&D Program of Hubei Province (2023BAB017, 2023BAB079), the National Natural Science Foundation of China (grants No.62072046, 62302181), HUST CSE-HongXin Joint Institute for Cyber Security, HUST CSE-FiberHome Joint Institute for Cyber Security, and the Xiaomi Young Talents Program.

\bibliographystyle{ACM-Reference-Format}
\bibliography{acmart}

\end{document}